\documentclass[a4paper,showpacs,superscriptaddress,10pt,nofootinbib]{revtex4}
\usepackage{amsmath,amssymb,bm,natbib}
\usepackage{epsfig}
\usepackage{graphicx}
\usepackage{multirow}
\usepackage{slashed}
\makeatletter

\newcommand{\Rmnum}[1]{\expandafter\@slowromancap\romannumeral #1@}
\makeatother
\begin{document}
\newcommand{\ttbar}{$t\overline{t}$\,\,}
\newcommand{\ee}{$e^+e^-$}
\newcommand{\nc}{\newcommand}
\nc{\jc}{\frac{1}{4}}  \nc{\sll}{S_{LL}}     \nc{\slr}{S_{LR}}
\nc{\srl}{S_{RL}}      \nc{\srr}{S_{RR}}     \nc{\vll}{V_{LL}}
\nc{\vlr}{V_{LR}}      \nc{\vrl}{V_{RL}}     \nc{\vrr}{V_{RR}}
\nc{\tll}{T_{LL}}      \nc{\tlrs}{T_{LR}}    \nc{\trl}{T_{RL}}
\nc{\trr}{T_{RR}}      \nc{\slld}{S_{LL}^D}  \nc{\slrd}{S_{LR}^D}
\nc{\srld}{S_{RL}^D}   \nc{\srrd}{S_{RR}^D}  \nc{\vlld}{V_{LL}^D}
\nc{\vlrd}{V_{LR}^D}   \nc{\vrld}{V_{RL}^D}  \nc{\vrrd}{V_{RR}^D}
\nc{\tlld}{T_{LL}^D}   \nc{\tlrd}{T_{LR}^D}  \nc{\trld}{T_{RL}^D}
\nc{\trrd}{T_{RR}^D}   \nc{\aqde}{\alpha_{qde}}
\nc{\alq}{\alpha_{\ell q}}        \nc{\alqp}{\alpha_{\ell q'}}
\nc{\alqt}{\alpha_{\ell q}^{(3)}} \nc{\alqtc}{\alpha_{\ell
q}^{(3)*}} \nc{\alqj}{\alpha_{\ell q}^{(1)}}
\nc{\alqjc}{\alpha_{\ell q}^{(1)*}} \nc{\aeu}{\alpha_{eu}}
\nc{\alu}{\alpha_{\ell u}} \nc{\aqe}{\alpha_{qe}}
\nc{\ber}{\begin{eqnarray*}} \nc{\enr}{\end{eqnarray*}}
\nc{\jmpb}{(1-\beta)/(1+\beta)} \nc{\wspR}{r}      \nc{\varx}{x}
\nc{\bt}{\beta}

\nc{\non}{\nonumber} \nc{\lspace}{\;\;\;\;\;\;\;\;\;\;}
\nc{\llspace}{\lspace \lspace}
\nc{\jnl}{\frac{1}{{\mit\Lambda}^2}} \nc{\jd}{\frac{1}{2}}
\def\eett{e^+e^-\to t\bar{t}}

\def\tt{t\bar{t}}
\nc{\bb}{\bibitem} \nc{\ra}{\rightarrow} \nc{\g}{\gamma}
\nc{\beq}{\begin{equation}} \nc{\eeq}{\end{equation}}

\def\dps{\displaystyle}

\title{
Top-spin analysis of new scalar and tensor 
interactions in
$e^{+} e^{-}$ collisions with beam polarization}
\author{{\bf B. Ananthanarayan }}
\author{{\bf Monalisa Patra}}
\affiliation{
Centre for High Energy Physics, 
Indian Institute of Science, 
Bangalore 560 012, India} 

\author{{\bf Saurabh D. Rindani}}

\affiliation{ 
Theoretical Physics Division,
Physical Research Laboratory,
Navrangpura, Ahmedabad 380 009, India}

%%ALIAS=ILCTDR=arXiv:0709.1893%%
%%ALIAS=BASDRPRD=hep-ph/0309260%%
%%ALIAS=BASDR1=hep-ph/0601199%%
%%ALIAS=BASDR2=arXiv:0805.2279%%
%%ALIAS=DR1=Dass:1975mj%%
%%ALIAS=DR2=Dass:1976wn%%
%%ALIAS=Grzadkowski=hep-ph/9511279%%
%%ALIAS=Hikasa=PHRVA.D33.3203%%
%%ALIAS=PolarizationReport=hep-ph/0507011%%
%%ALIAS=BW=NUPHA.B268.621%%
%%ALIAS=BR=IMPAE.A6.2707%%
%%ALIAS=BurgessRobinson=Burgess:1990ba%%
\begin{abstract}
We utilize top polarization in 
the process $e^{+} e^{-}\rightarrow t\overline{t}$ 
at the ILC with transverse beam polarization to probe 
interactions of the scalar and tensor type beyond the standard model
and to
disentangle their individual contributions.
90\% confidence level limits on the interactions with realistic integrated 
luminosity are presented and are found
to improve by an order of magnitude compared to the case
when the spin of the top quark is not measured. Sensitivities of 
the order of a few times $10^{-3}\, {\rm TeV}^{-2}$ for
real and imaginary parts of both scalar and tensor couplings 
at $\sqrt{s}$=500 and 800 {\rm GeV} with an integrated luminosity
of 500 ${\rm fb}^{-1}$ and completely polarized beams is shown to be
possible. 
A powerful model-independent framework for
inclusive measurements is employed to describe the 
spin-momentum correlations and their C, P and T properties is
presented in a technical appendix. 
\end{abstract}

\pacs{11.30 Er,13.66 Jn,13.88.+e}

\maketitle

\section{Introduction}
At the planned ILC~\cite{ILCTDR} $e^{+} e^{-}\rightarrow$ \ttbar  is 
a process that will be studied at great precision to validate the
standard model (SM) and to look for deviations from it.  
The process is of continued current interest, 
see e.g., ref.~\cite{arXiv:1005.1756} and references therein.
The availability of beam polarization will significantly
enhance the sensitivity to new physics(NP) provided the electron and
positron beams have transverse polarization (TP) or longitudinal
polarization, each complementing the other, with
distinct prospects of obtaining very high
degree of polarization for both beams~\cite{PolarizationReport}.

One fruitful approach is to undertake
a model independent analysis which may be performed 
by introducing higher dimensional operators consistent with gauge invariance,
for an important early paper on the subject, see ref.~\cite{BW}.
In the context of top pair production, the relevant 
higher dimensional operators are listed in ref.~\cite{Grzadkowski}
and references therein.
In this work we will confine ourselves to NP 
associated only with scalar and tensor type operators
which cannot be probed at linear order unless TP is available.  They
are parametrized in terms of operators denoted by
$S_{RR}$ and $T_{RR}$ in ref.~\cite{Grzadkowski}
which will hereafter be denoted by $S$ and $T$ respectively.
In a recent work~\cite{BASDRPRD} it was shown that in the presence of TP 
if only cross sections were to be measured, azimuthal asymmetries would 
involve the linear combination given by
\begin{equation}\label{eqim}
 S+\frac{2c^{t}_{A}c^{e}_{V}}{c^{t}_{V}c^{e}_{A}}T,
\end{equation}
where $c^f_V$ and $c^f_A, \, f=e,t$ are the vector and
axial-vector couplings of the $Z$ to
the electron and the top-quark respectively and will be explicitly
given later.
Therefore, it becomes important to pose the question of how these
operators can be disentangled.  The aim of the present work is
to address this question and to demonstrate that measurement of
the top-quark spin can indeed allow one to disentangle them.
Indeed, this work is also motivated by the fact that it may
be possible now to measure the top spin accurately, see, e.g.,
ref.~\cite{Brandenburg:1999ss}.  Also, it has been recently
pointed out
that top polarization can be measured reliably from decay charged-lepton
angular distributions without errors arising from the $tbW$ 
couplings~\cite{hep-ph/0605100}.

To meet the ends described above, 
we have evaluated the beam polarization dependent differential
cross sections and examined the polarization of $t$ or $\bar t$.
We have checked our 
{\it ab initio} evaluation of the cross sections 
with the results from the explicit helicity
amplitudes provided by Grzadkowski~\cite{Grzadkowski} 
combined with the general framework for inclusion of
TP provided in ref.~\cite{Hikasa}.  We have also carried out
independent check on the helicity amplitudes.  

We will finally explore the reach of the ILC by defining suitable
observables and asymmetries. 
Turning to the numerical implications of our work, 
we find surprisingly that the top-spin resolution allows us to
probe the NP at a level an order of magnitude better
than the reach reported in~\cite{BASDRPRD}.  
In spirit, therefore,
this work is a natural completion of the prior work on $t\bar{t}$ production
and that of our work on inclusive processes with only momentum structure
functions, thereby providing a systematic contribution to the
physics programme at the ILC with polarized beams.

In order to understand the formal structure of 
the terms in the distributions,
we consider them in detail and 
isolate the spin-independent and spin-dependent parts 
of the cross section and present the results
in a technical appendix.  The former have been
interpreted in terms of momentum structure functions
arising in the treatment of the inclusive process framework of the 
type considered recently by us in~\cite{BASDR1},
which relies on an analysis of:
\begin{equation}\label{process1}
e^++e^-\rightarrow h(p)+X 
\end{equation}
where $h$ is the observed final state particle whose momentum $p$
is measured, and $X$ here and elsewhere
refers to an inclusive state.
Furthermore, it was shown  
even an exclusive process such as \ttbar production can be
included in the framework of the general inclusive process~\cite{BASDR1}.
Also in ref.~\cite{BASDR2} the two-particle inclusive process
\begin{equation}\label{process2}
e^++e^-\rightarrow h_1(p_1)+h_2(p_2)+X 
\end{equation}
where $h_{1,2}$ are the observed final state particles whose momenta
$p_{1,2}$ are measured was considered.  The above can be generalized 
as in the work of~\cite{DR2}
to
\begin{equation}\label{process3}
e^++e^-\rightarrow h(p,s)+X 
\end{equation}
where $p$, $s$ stand respectively for the momentum and spin of the observed 
particle $h$ in the final state.
However this was not performed explicitly in~\cite{BASDR2}.  
The present work gives us an opportunity to do so in the
context of the process at hand, thus providing a
concrete illustration.
The C, P and T properties of the
operators will be considered, and those of the structure functions
separately so that the two frameworks can be related to one another.
Furthermore, one may obtain insights into discrete symmetries of NP
contributions from the nature of spin-momentum correlations.
Important insights based on general
inclusive processes have enriched the analysis of processes such as 
$Z\gamma$ production~\cite{hep-ph/0404106,hep-ph/0410084,hep-ph/0507037},
$Zh$ production~\cite{hep-ph/0605298,arXiv:0709.2591}.

The scheme of this paper is as follows:
In Sect.~\ref{formalism} 
we will recall the main features of the framework where
NP is introduced in terms of effective four-Fermi interactions.
In Sect.~\ref{distributions}  
we will consider the cross sections for \ee$\rightarrow$\ttbar
in the presence of four-Fermi NP interactions. 
In Sect.~\ref{numerical}  
we consider applications using realistic luminosity and degrees
of polarization at typical ILC energies of 500 and 800 {\rm GeV} and obtain
the 90\% confidence level (C.L.) limits that can be placed on the NP 
operators by constructing suitable asymmetries.   We note here that
this is in an idealized situation, where a realistic measurement would
produce somewhat weaker conclusions which we have not attempted to
assess here.
In Sect.~\ref{polarization} we turn to the important question of
realizing the proposal in terms of an actual experimental measurement.
In Sect.~\ref{conclusions}  
we provide a thorough discussion of the various features
emerging from our investigations and present our conclusions.
In the Appendix
we interpret the results obtained in the preceding section
in terms of momentum and spin structure functions.
The corresponding properties under the discrete symmetries C, P and T
are discussed.

\section{Effective Operators and four-Fermi interaction}\label{formalism}
The theoretical framework that we consider is one of the SM augmented with
four-Fermi interactions that capture the effect of all the NP.
In particular, for the process $\eett$, the
tree level operators which will contribute are 
(see ref.~\cite{BASDRPRD} and references therein).
\begin{equation}\label{operators}
\begin{array}{lcl}
{\cal O}^{(1)}_{\ell q}&\!\!=&\!\!\dps\frac12
(\bar{\ell}\gamma_{\mu}\ell)(\bar{q}\gamma^{\mu}q),\\ && \\ {\cal
O}^{(3)}_{\ell q}&\!\!=&\!\!\dps\frac12
(\bar{\ell}\gamma_{\mu}\tau^I\ell) (\bar{q}\gamma^{\mu}\tau^Iq),\\
\vspace*{-0.3cm} & &  \\ \vspace*{-0.3cm} {\cal O}_{eu}&\!\!
=&\!\!\dps\frac12 (\bar{e}\gamma_{\mu}e)(\bar{u}\gamma_{\mu}u),
\\ & &  \\ {\cal O}_{\ell u}&\!\!=&\!\!(\bar{\ell}u)(\bar{u}\ell),
\\ {\cal O}_{qe}&\!\!=&\!\!(\bar{q}e)(\bar{e}q),\\ {\cal O}_{\ell
q}&\!\!=&\!\!(\bar{\ell}e)\epsilon (\bar{q}u), \\ {\cal O}_{\ell
q'}&\!\!=&\!\!(\bar{\ell}u)\epsilon (\bar{q}e),
\end{array}
\end{equation}
where $l,q$ denote respectively the left-handed electroweak $SU(2)$
lepton and quark doublets, and $e$ and $u$ denote $SU(2)$ singlet
charged-lepton and up-quark right-handed fields. $\tau^I$ $(I=1,2,3)$ are
the Pauli matrices, and $\epsilon$ is the $2\times 2$ 
anti-symmetric matrix, $\epsilon_{12}=-\epsilon_{21}=1$, and
generation indices are suppressed.
Given the above operators, the Lagrangian which we will
use is:
\begin{equation}\label{lag}
{\cal L}={\cal L}^{S\!M}+
\frac{1}{{\mit\Lambda}^2}\sum_i(\:\alpha_i{\cal O}_i+{\rm
h.c.}\:),
\end{equation}
where $\alpha$'s are the coefficients which parametrize
non-standard interactions. 
The NP four-Fermi operators contained in the Lagrangian
after Fierz transformation takes the form
\begin{equation}\label{lag4f}
{\cal L}^{4F}
 =\sum_{i,j=L,R}\Bigl[\:S_{ij}(\bar{e}P_ie)(\bar{t}P_jt)
 +T_{ij}
 (\bar{e}\frac{\sigma_{\mu\nu}}{\sqrt{2}}P_ie)
(\bar{t}\frac{\sigma^{\mu\nu}}{\sqrt{2}}P_jt)\:\Bigr]
\end{equation}
with the coefficients satisfying the following constraints:
\begin{eqnarray}
S\equiv S_{RR}=S^{*}_{LL},\ \ \ S_{LR}=S_{RL}=0,% \nonumber \\ &&
 T\equiv T_{RR}=T^{*}_{LL},\ \ \
T_{LR}=T_{RL}=0.
\end{eqnarray}
In (\ref{lag4f}), $P_{L,R}$ are respectively the left- and right-chirality
projection matrices and the correspondence between the
$\alpha_i$ and the $S(T)_{ij}$ may be read off from~\cite{Grzadkowski}. 
It may be recalled here that a significant discussion was provided
in ref.~\cite{BASDRPRD} on the scale of the operators that can arise
from considerations of naturalness as well as constraints arising
from such considerations as the electron electric and magnetic
dipole moments.

\section{Distributions in the presence of polarization}\label{distributions} 
We consider the process $e^+e^-\rightarrow t \overline
t$ for the cases when the spin of the top quark is measured and
the spins of the $\overline{t}$ are summed over, and vice-versa.
We wish to examine the CP-violating as well as
conserving contributions in the interference
of the SM amplitude with the scalar and tensor four-Fermi
amplitudes. We will take the electron TP to be 100\% and along the
positive or negative $x$ axis, and the positron polarization to be
100\%, parallel or anti-parallel to the electron polarization.
The $z$ axis is chosen along the direction of the $e^-$. The
differential cross sections for $e^+e^-\rightarrow t \overline t$,
with the superscripts denoting the respective signs of the $e^-$
and $e^+$ TP retaining the new couplings to linear order only are:
\begin{eqnarray}\label{diffcspp1}
{\displaystyle \frac{d\sigma^{\pm\pm}}{d\Omega}}&=&
{\displaystyle \frac{d\sigma^{\pm\pm}_{SM}}{d\Omega}}\pm
{\displaystyle \frac{h\alpha\beta}{\pi s^2}}[\frac{s^{3/2}}{2} m_t \sin\theta
({\rm Re}T \cos\phi-\frac{1}{2}z^\prime({\rm Im}S-
2\beta {\rm Im}T\cos\theta)\sin\phi)] \nonumber \\
&& \pm \displaystyle{\frac{\alpha\beta}{4\pi s(s-M_{z}^{2})}}
\left[-3h\beta m_t c^t_A c^e_A \cos\theta\sin\theta\cos\phi s^{3/2}
{\rm Re} T \right. \nonumber \\
&& \left. +{\displaystyle \frac{3}{2}}m_t s^{3/2}\sin \theta 
\left\{(\beta c^e_A c^t_V {\rm Re} S 
 +2 c^e_V(\beta c^t_A-h c^t_V){\rm Re} T)\cos\phi \right. \right. \nonumber \\
&& \left. \left. +{\displaystyle z^\prime h c^e_V c^t_V({\rm Im} S -
2 \beta {\rm Im} T \cos\theta)\sin \phi} \right\}\right]
\end{eqnarray}
and
\begin{eqnarray}\label{diffcspp2}
\displaystyle{\frac{d\sigma^{\pm\mp}}{d\Omega}}&=&
\displaystyle{\frac{d\sigma^{\pm\mp}_{SM}}{d\Omega}}\mp
\displaystyle{\frac{h\alpha\beta}{\pi s^2}}[\frac {s^{3/2}}{2} m_t \sin\theta
({\rm Im} T \sin\phi-\frac{1}{2}z^\prime({\rm Re}
 S-2\beta {\rm Re} T\cos\theta)\cos\phi)] \nonumber \\
&& \pm \displaystyle{\frac{\alpha \beta}
{4\pi s(s-M_{z}^{2})}}
\left[3h\beta m_t c^t_A c^e_A \cos\theta\sin\theta\sin\phi s^{3/2}
{\rm Im} T \right. \nonumber \\
&& \left. -\displaystyle{\frac{3}{2}}m_t s^{3/2}\sin \theta 
\{(\beta c^e_A c^t_V {\rm Im} S + 2 c^e_V(\beta c^t_A-h c^t_V){\rm Im} T)\sin\phi \right. \nonumber \\
&& \left. + \displaystyle{z^\prime h c^e_V c^t_V({\rm Re} S -2 \beta 
{\rm Re} T \cos\theta)\cos \phi} \}\right]
\end{eqnarray}
where
\begin{eqnarray}\label{diffcsSM}
\displaystyle{\frac{d\sigma^{+\pm}_{SM}}{d\Omega}}&=&
\frac{d\sigma^{-\mp}_{SM}}{d\Omega}\non\\
&=&\frac{3\alpha^2\beta}{4s}\left[\frac{4}{9}
\{\frac{1}{2}(1+\cos^2\theta)+\frac{2 m_t^2}{s}\sin^2\theta\pm\frac{1}{2}
\beta^2\sin^2\theta\cos2\phi\}\right.\non\\
&&\left. -\frac{s}{s-M_Z^2}\frac{4}{3}\left 
\{\frac{1}{2}c^e_V(c^t_V-h\beta c^t_A)(1+\cos^2\theta)+
\frac{2m_t^2}{s}c^e_Vc^t_V\sin^2\theta\right. \right. \non\\
&& \left. \left.+c^e_A(\beta c^t_A-hc^t_V)\cos\theta\pm
\frac{1}{2}\beta c^e_V(\beta c^t_V-hc^t_A)
\sin^2\theta\cos2\phi\right\}\right.\non \\
&& \left.+\frac{s^2}{(s-M_Z^2)^2}\left 
\{\frac{1}{2}(c^{e^2}_A+c^{e^2}_V)((c^t_V-h\beta c^t_A)^2
(1+\cos^2\theta)+\frac{4m_t^2}{s}c^{t^2}_V\sin^2\theta)\right. \right. \non\\
&&\left. \left. -2hc^e_Ac^e_V(c^{t^2}_V+\beta^2 c^{t^2}_A)\cos\theta
+4\beta\cos\theta c^e_Ac^e_Vc^t_Ac^t_V \right. \right. \non\\
&&\left. \left. \pm\frac{1}{2}(c^{e^2}_V-c^{e^2}_A)
(\beta^2(c^{t^2}_A+c^{t^2}_V)-2h\beta c^t_Ac^t_V)\sin^2\theta\cos2\phi \right \}\right]
\end{eqnarray}
with
$\beta = \sqrt{1-4m_t^2/s}$, and 
$c_V^i$, $c_A^i$ as the couplings of $Z$ to $e^-e^+$ and $t
\overline t$.  Explicitly the couplings are:
\begin{eqnarray}
\displaystyle c^e_V=\frac{1}{2\sin\theta_W \cos\theta_W}
\left(-\frac{1}{2}+2\sin^2\theta_W \right), &
\displaystyle c^e_A=-\frac{1}{4\sin\theta_W \cos\theta_W}, \nonumber \\
\displaystyle c^t_V=\frac{1}{2\sin\theta_W \cos\theta_W}
\left(\frac{1}{2}-\frac{4}{3}\sin^2\theta_W \right), & 
\displaystyle c^t_A=\frac{1}{4\sin\theta_W \cos\theta_W}.
\end{eqnarray}
In the above, $h$ stands for the helicity of the top quark when the
spin of the $\overline{t}$ is summed over, and for the negative of
the helicity of the $\overline{t}$ when the spin of the top quark
is summed over. 
The following may be noted:
(a) the part of the distribution independent
of the final state helicity was already given in ref.~\cite{BASDRPRD} and
that a sign error in the NP contributions therein is corrected here,
and (b) $z'$ appears only in the NP contributions and 
is $+1$ for the top quark and $-1$ for
$\overline{t}$.

In order to render these expressions useful for ILC applications,
and to disentangle the separate NP effects, we will define asymmetries
that will isolate their individual contributions.
These will be employed to obtain 90\% confidence level limits
on the NP couplings with realistic integrated luminosities
in the absence of any signal at the ILC.

The explicit expressions in terms of the
laboratory observables such as the momenta, polar and
azimuthal angles, accompanying the helicity independent
and helicity dependent parts require a detailed discussion.  
Since the expressions above are quite involved, in order to get a better
insight into the nature of the spin-momentum correlations and
spin-spin correlations, in the Appendix, we will consider
a general framework first developed for a general inclusive
process.  This will enable us to interpret the angular correlations
in terms of the vectorial quantities that define the process.
Furthermore, it will also enable us to obtain insights into the
symmetry properties of the correlations under the discrete symmetries
C, P and T and study the consequences of the CPT theorem.

\section{Extraction of new physics}\label{numerical}
In this section we now address the question of isolating the
contributions from the NP by constructing suitable asymmetries.
Clever choices can lead to asymmetries receiving contributions
from only one of them, while the others cancel out due to 
integrations over polar as well as azimuthal angles.  Whereas
in the helicity independent case it was impossible to disentangle
the scalar and tensor contributions, now the rich structure
of the helicity dependent parts allows us to meet the objective
that we have set out.

One may also ask to what extent this can be achieved if only
one electron and positron spin configuration is available.  
Even in this case it is possible to isolate the NP contribution
term by term.  
Finally, we explore the situation when all spin configurations
are available.

 In this section we have isolated the contributions
coming from NP using different asymmetries.
A thorough numerical analysis has been done to place constraints
on the anomalous couplings.
\subsection{Integrated Asymmetries}
For the case of angular distribution with transversely polarized beams,
there is a dependence on the azimuthal angle. Compared to the unpolarized 
case there are various terms with combinations such as 
$\sin\theta\cos\phi$, $\sin\theta\sin\phi$,
$\sin\theta\cos\theta\sin\phi$, $\sin\theta\cos\theta\cos\phi$.
We define below different azimuthal asymmetries which are used 
to isolate the couplings. The generic forms of the asymmetries,
for the moment
suppressing the beam polarizations, are:
\begin{eqnarray}
&& A_1(\theta)=\frac{1}{\sigma^{SM}(\theta)}\biggl[\int^\pi_0\frac{
d\sigma_{NP}}{d\Omega}\,d\phi-\int^{2\pi}_\pi\frac{d\sigma_{NP}}{d\Omega}\,
d\phi\biggr]\label{asym1} \\
&& A_2(\theta)=\frac{1}{\sigma^{SM}(\theta)}\biggl[\int^{\frac{\pi}{2}}_{
0}\frac{d\sigma_{NP}}{d\Omega}\,d\phi-\int^{\frac{3\pi}{2}}_{
\frac{\pi}{2}}\frac{d\sigma_{NP}}{d\Omega}\,d\phi+\int^{2\pi}_{\frac{3\pi}{2}}
\frac{d\sigma_{NP}}{d\Omega}\,d\phi\biggr] \label{asym2}
\end{eqnarray}
where
\begin{equation}
\frac{d\sigma_{NP}}{d\Omega}={\left.\frac{d\sigma}{d\Omega}\right|}_{h=1}-
{\left.\frac{d\sigma}{d\Omega}\right|}_{h=-1}
\end{equation}
and
\begin{equation}
\sigma^{SM}(\theta)=\biggl[\int^{2 \pi}_0\frac{d\sigma_{SM}}{d\Omega}\,d\phi
\biggr].
\end{equation}

The above asymmetries amount to  correlations
between the spin direction of the top quark and its production angle.
Expressed in a different language, they are azimuthal asymmetries calculated 
for the top polarization dependent part of the differential
cross section.

\begin{eqnarray}
\text{Case 1:}\quad \left\{
\begin{array}{l l l}
&&A_1^{+-}(\theta)={\displaystyle\frac{1}{\sigma^{SM}(\theta)}
\frac{2m_t\alpha\beta\sin\theta}{\pi\sqrt{s}}}
\biggl[-2+\frac{3s}{s-M_Z^2}(c^e_Vc^t_V+\beta c^e_A c^t_A\cos\theta)\biggr]
{\rm Im} T \label{eqnt1}\\
&& A_2^{+-}(\theta)={\displaystyle\frac{1}{\sigma^{SM}(\theta)}
\frac{2m_t\alpha\beta\sin\theta}{\pi\sqrt{s}}}
\biggl[-1+\frac{3s}{2(s-M_Z^2)}c^e_Vc^t_V\biggr]
\bigl(2{\rm Re} T\beta\cos\theta - {\rm Re} S\bigr) \label{eqnt2}\\
\end{array} \right.
\end{eqnarray}
\begin{eqnarray}
\text{Case 2:}\quad \left\{ 
\begin{array}{l l l}
&& A_1^{++}(\theta)={\displaystyle\frac{1}{\sigma^{SM}(\theta)}
\frac{2m_t\alpha\beta\sin\theta}{\pi\sqrt{s}}}
\biggl[-1+\frac{3s}{2(s-M_Z^2)}c^e_Vc^t_V\biggr]
\bigl({\rm Im} S-2 {\rm Im} T\beta\cos\theta \bigr)\label{eqnt3}\\
&& A_2^{++}(\theta)={\displaystyle\frac{1}{\sigma^{SM}(\theta)}
\frac{2m_t\alpha\beta\sin\theta}{\pi\sqrt{s}}}
\biggl[2-\frac{3s}{s-M_Z^2}(c^e_Vc^t_V+\beta c^e_A c^t_A\cos\theta)\biggr]
{\rm Re} T \label{eqnt4}\\
\end{array} \right.
\end{eqnarray}

The choice of our asymmetries can be justified by taking a 
close look at the expressions above. Confining ourselves to Case 1,
it is seen that $A_1^{+-}(\theta)$ depends solely on ${\rm Im} T$,
whereas $A_2^{+-}(\theta)$ is proportional to both ${\rm Re} T$ and
${\rm Re} S$. Similarly for Case 2 the coupling ${\rm Re} T$
can be isolated from $A_2^{++}(\theta)$, whereas
$A_1^{++}(\theta)$ is proportional to both ${\rm Im} S$ and
${\rm Im} T$. Before proceeding further 
we would like to point out that, when the final state helicity
is summed over, only one asymmetry is non-zero for each beam polarization combination~\cite{BASDRPRD}:
\begin{eqnarray}
&\text{Case 1:} \quad & \hat{A}_1^{+-}(\theta)=-{\displaystyle\frac{1}{\sigma^{SM}(\theta)}
\frac{2m_t\alpha\beta\sin\theta}{\pi\sqrt{s}}}
\biggl[{\displaystyle\frac{3}{2}\frac{s}{s-m_Z^2}c^e_Ac^t_V\beta {\rm Im} 
(S+\frac{2c^{t}_{A}c^{e}_{V}}{c^{t}_{V}c^{e}_{A}}T)}
\biggr]  \\
& \text{Case 2:} \quad & \hat{A}_2^{++}(\theta)={\displaystyle\frac{1}{\sigma^{SM}(\theta)}\frac{2m_t\alpha\beta\sin\theta}{\pi\sqrt{s}}}\biggl[{\displaystyle\frac{3}{2}\frac{s}{s-m_Z^2}c^e_Ac^t_V\beta {\rm Re} 
(S+\frac{2c^{t}_{A}c^{e}_{V}}{c^{t}_{V}c^{e}_{A}}T)
}\biggr] 
\end{eqnarray}

where 

\begin{eqnarray}
&& \hat{A}^{+-}_1(\theta)=\frac{1}{\sigma^{SM}(\theta)}\biggl[\int^\pi_0\frac{
d\sigma_{tot}^{+-}}{d\Omega}\,d\phi-\int^{2\pi}_\pi\frac{d\sigma_{tot}^{+-}}{d\Omega}\,
d\phi\biggr]\label{asym1p} \\
&& \hat{A}^{++}_2(\theta)=\frac{1}{\sigma^{SM}(\theta)}\biggl[\int^{\frac{\pi}{2}}_{
0}\frac{d\sigma_{tot}^{++}}{d\Omega}\,d\phi-\int^{\frac{3\pi}{2}}_{
\frac{\pi}{2}}\frac{d\sigma_{tot}^{++}}{d\Omega}\,d\phi+\int^{2\pi}_{\frac{3\pi}{2}}
\frac{d\sigma_{tot}^{++}}{d\Omega}\,d\phi\biggr] \label{asym2p}
\end{eqnarray}
and
\begin{equation}
\frac{d\sigma_{tot}^{+\mp}}{d\Omega}={\left.\frac{d\sigma^{+\mp}}{d\Omega}\right|}_{h=1}+
{\left.\frac{d\sigma^{+\mp}}{d\Omega}\right|}_{h=-1}
\end{equation}

The $\theta$-integrated version of the asymmetries in eqs.
(\ref{asym1}), (\ref{asym2}) is:
\begin{equation}
A_1(\theta_0)=\frac{1}{\Delta\sigma^{SM}(\theta_0)}\biggl[\int^{\cos\theta_0}_{-\cos\theta_0}\biggl(\int^\pi_0\frac{d\sigma_{NP}}{d\Omega}\,d\phi-\int^{2\pi}_\pi\frac{d\sigma_{NP}}{d\Omega}\,d\phi\biggr)\,d\cos\theta\biggr]
\end{equation}
where 
\begin{equation}
\Delta\sigma^{SM}(\theta_0)=\biggl[{\displaystyle\int^{\cos\theta_0}_{
-\cos\theta_0}\biggl(\int^{2 \pi}_0\frac{d\sigma_{SM}}{d\Omega}
\,d\phi \biggr)\,d\cos\theta}\biggr]
\end{equation} 
is independent of the transverse beam polarizations. $A_2(\theta_0)$ can
be defined analogously to $A_1(\theta_0)$ above.
A cut-off on $\theta$ has been introduced above,
for a practical reason to stay away from the beam pipe.
The asymmetries with the given limit on $\theta$, $\theta_0< \theta < 
\pi-\theta_0$,
can be easily obtained.
After the introduction of cut-off the terms
proportional to $\cos\theta$ vanish. Limiting ourselves
to Case 1 we see that $A_1^{+-}(\theta_0)$ depends on ${\rm Im} T$,
and  $A_2^{+-}(\theta_0)$ depends on ${\rm Re} S$. It is seen that the coefficient
of ${\rm Im} T$ in $A_1^{+-}(\theta_0)$ is twice that of ${\rm Re}
S$ in  $A_2^{+-}(\theta_0)$.

Continuing our analysis further, we note that we can determine 
only two of the four couplings using either $++$ or $+-$ polarizations.
For the determination of all of them both polarization combinations 
have to be used.
Restricting ourselves to the possibility that only one 
polarization combination is available,
we can consider an additional asymmetry which combines a forward-backward asymmetry
with an additional asymmetry in $\phi$:
\begin{small}
\begin{eqnarray}
A_1^{FB}(\theta_0)&=& \frac{1}{\Delta\sigma^{SM}(\theta_0)} \nonumber \\
&& \biggl[\int^{\cos\theta_0}_{0}\,d\cos\theta 
\biggl(\int^\pi_0\frac{d\sigma_{NP}}
{d\Omega}\,d\phi-\int^{2\pi}_\pi\frac{d\sigma_{NP}}
{d\Omega}\,d\phi\biggr)-\int^{0}_{-\cos\theta_0}\,d\cos\theta 
\biggl(\int^\pi_0\frac{d\sigma_{NP}}
{d\Omega}\,d\phi-\int^{2\pi}_\pi\frac{d\sigma_{NP}}
{d\Omega}\,d\phi\biggr)\biggr]
\end{eqnarray}
\end{small}
$A_2^{FB}(\theta_0)$ can be defined in an analogous way.
These are easily evaluated for the $+-$ case:
\begin{eqnarray}
&& A_1^{FB}(\theta_0)=\frac{1}{\Delta\sigma^{SM}(\theta_0)}\frac{4m_t\alpha\beta}{\pi\sqrt{s}}\biggl[\frac{s}{s-M_Z^2}\beta c^e_Ac^t_A\biggr](1-\sin ^3\theta_0) {\rm Im} T  \\
&& A_2^{FB}(\theta_0)=\frac{1}{\Delta\sigma^{SM}(\theta_0)}\frac{4m_t\alpha\beta^2}{3\pi\sqrt{s}}\biggl[-2 + 3\frac{s}{s-M_Z^2} c^e_Vc^t_V\biggr](1-\sin^3\theta_0){\rm Re} T 
\end{eqnarray}
The above expressions show $A_1^{FB}(\theta_0)$
depends on ${\rm Im} T$, and $A_1^{FB}(\theta_0)$ is proportional to 
${\rm Re} T$. Thus all the couplings 
available for a single polarization combination
can be isolated using the above asymmetries. 

We have done a thorough numerical analysis
in the next sub section for $+-$ case. 
The $++$ case can be treated analogously.

\subsection{Numerical results}

We have calculated the asymmetries
under the ideal condition of 100\% beam polarization for $e^-$
as well as $e^+$ at $\sqrt{s}$=500 {\rm GeV} and 800 {\rm GeV} respectively for an integrated
luminosity of 500 ${\rm fb}^{-1}$.
An explicit calculation has been done for
the $(+-)$ case. 

\begin{figure}[htb]
\begin{center}
\includegraphics[width=6.5cm,height=5 cm]{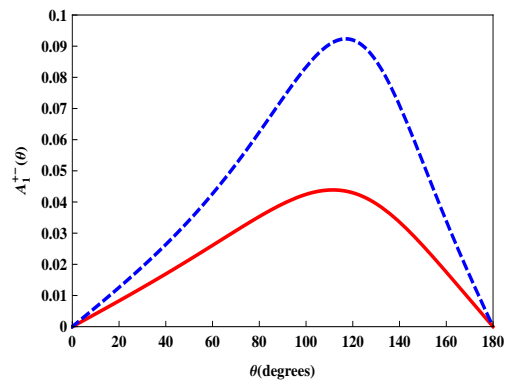}
\caption{$A_1^{+-}(\theta)$ as a function of $\theta$ for a
value of {\rm Im}$T$=0.01 ${\rm TeV}^{-2}$ at $\sqrt{s}=500$ 
{\rm GeV} [Red-Solid] and $\sqrt{s}=800$ {\rm GeV} [Blue-Dashed]. }
\label{fig:F1}
\end{center}
\end{figure}
\begin{figure}[htb]
\includegraphics[width=6.5cm,height=5.1 cm]{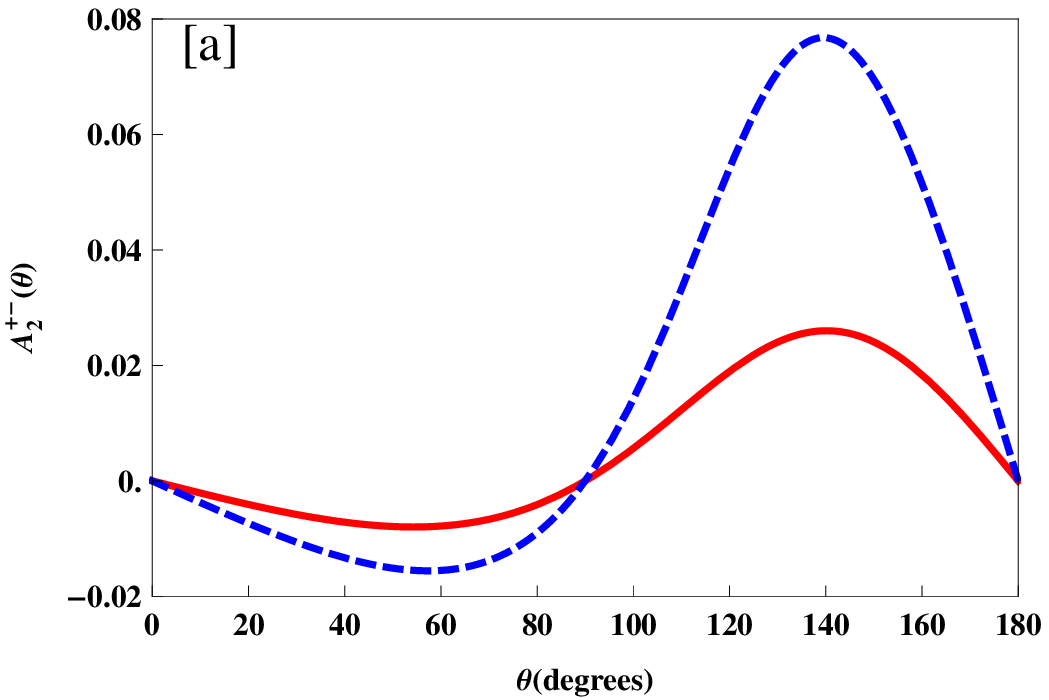}
\hspace{0.2cm}
\includegraphics[width=6.5 cm,height=5 cm]{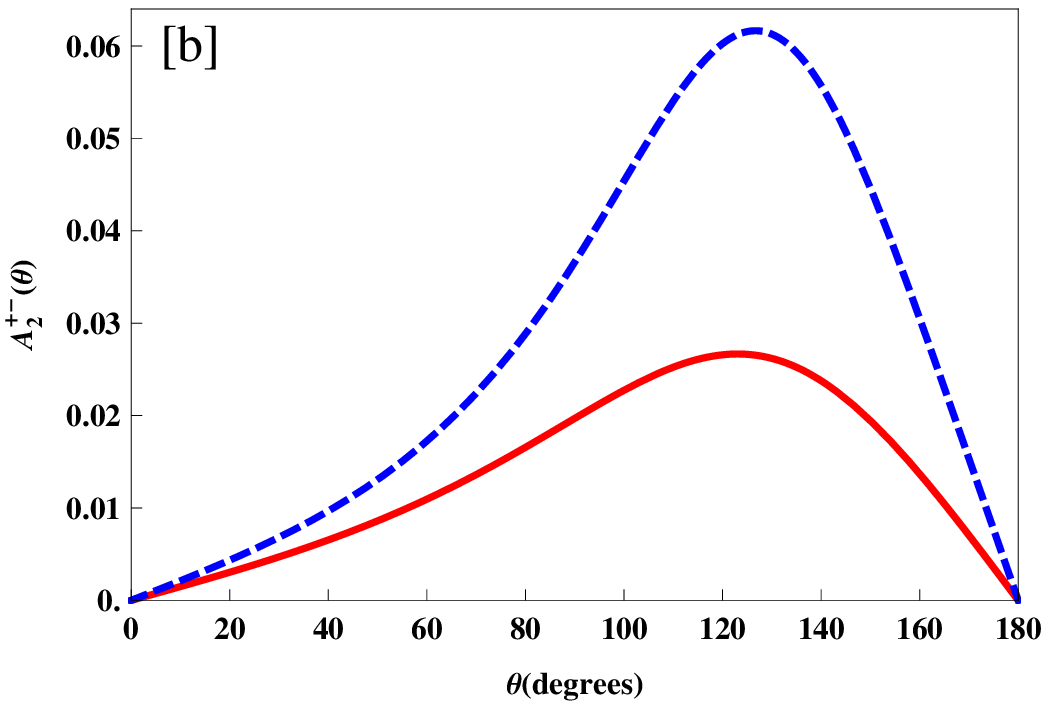}
\caption{$A_2^{+-}(\theta)$ as a function of $\theta$ for [a] {\rm Re}$T$=0.01
${\rm TeV}^{-2}$, {\rm Re}$S$=0,
[b] {\rm Re}$S$=0.01 ${\rm TeV}^{-2}$, {\rm Re}$T$=0 at $\sqrt{s}=500$
{\rm GeV} [Red-Solid] and $\sqrt{s}=800$ {\rm GeV} [Blue-Dashed]. }
\label{fig:F23}
\end{figure}

We first plot $A_1^{+-}(\theta)$ and $A_2^{+-}(\theta)$ in Fig.~\ref{fig:F1}
and Fig.~\ref{fig:F23} as a function of $\theta$ for different
centre-of-mass (c.m.) energies. In Fig.~\ref{fig:F1} $A_1^{+-}(\theta)$ is plotted
for {\rm Im}$T$ =0.01 ${\rm TeV}^{-2}$. Fig.~\ref{fig:F23} shows $A_2^{+-}(\theta)$ which 
is a function of the anomalous couplings {\rm Re}$T$ and {\rm Re}$S$, plotted
as a function of $\theta$ with one coupling taken to be non zero at a time. 
All the asymmetries here not only 
vanish for 
$0\,^{\circ}$ and $180\,^{\circ}$ because they are proportional to
$\sin\theta$, but $A_2^{+-}(\theta)$ for
{\rm Re}$T$
also vanishes for $90\,^{\circ}$ as it is additionally proportional  to 
$\cos\theta$. The asymmetry $A_1^{+-}(\theta)$ can be as high as 9\% in the future
colliders for $\sqrt{s}$=800  GeV whereas $A_2^{+-}(\theta)$ can
attain a maximum of 8\% for  {\rm Re}$T$ and 6 \% for  {\rm Re}$S$ at the same
c.m. energy, for the chosen values of the parameters as 0.01 ${\rm TeV}^{-2}$ each . 
%The sensitivity
%range of the asymmetries are about 6 \% to 9 \% in the future colliders.

\begin{figure}[htb]
\includegraphics[width=6.5cm,height=5 cm]{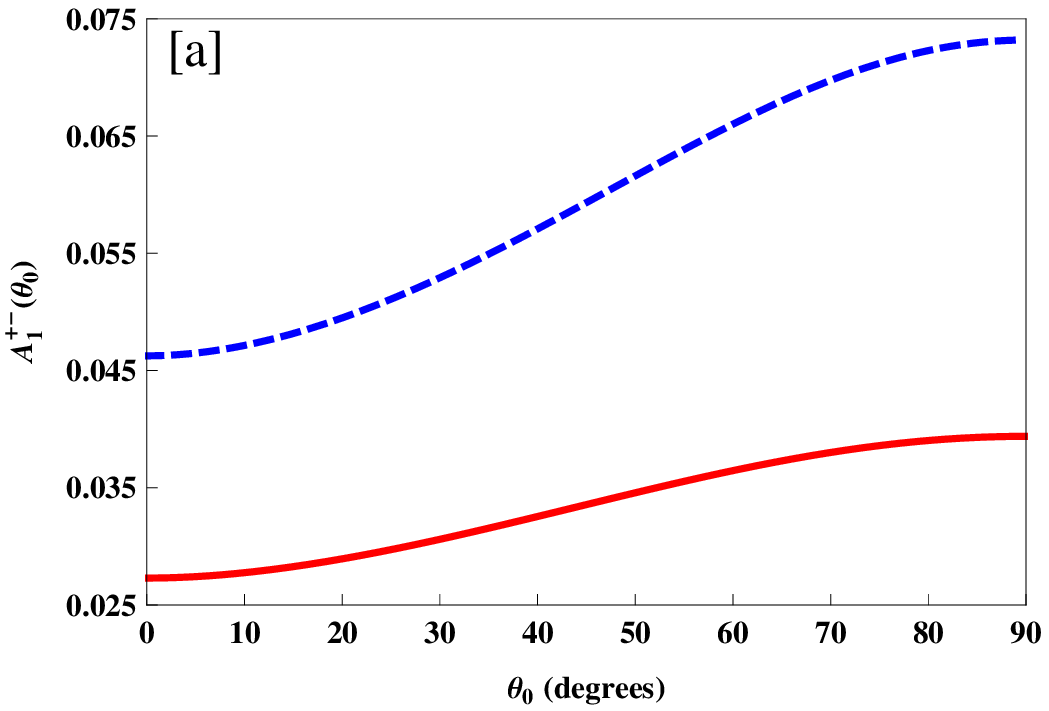}
\hspace{0.2cm}
\includegraphics[width=6.5 cm,height=5 cm]{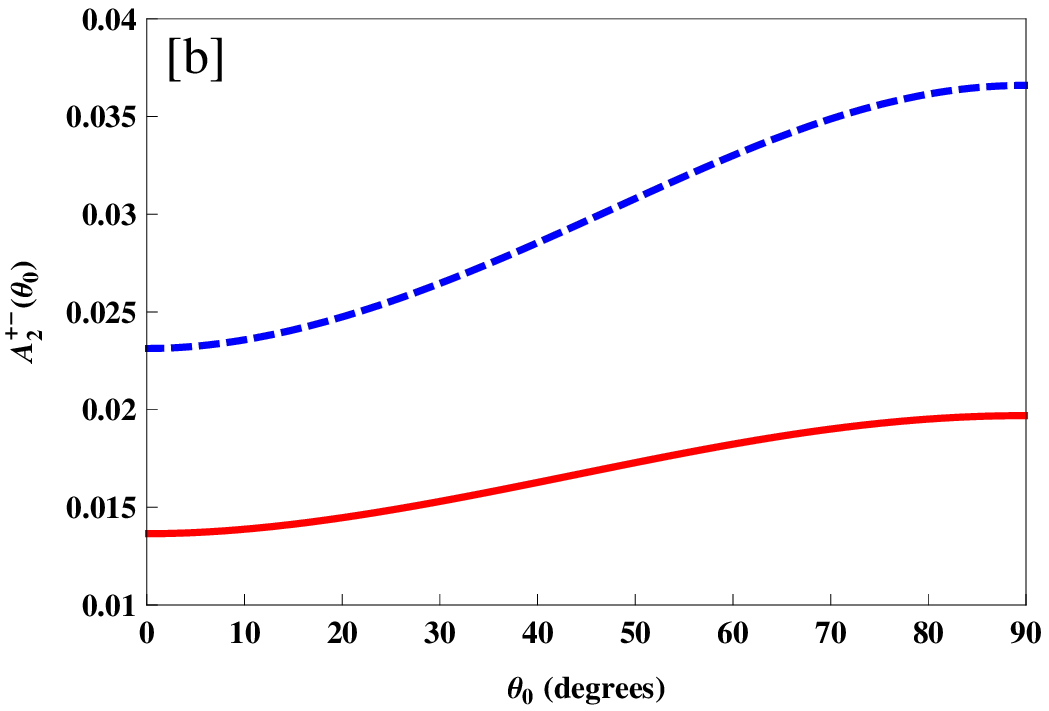}
\caption{The asymmetries as a function of $\theta_0$
for $\sqrt{s}$=500 {\rm GeV} [Red-Solid] and $\sqrt{s}$=800 {\rm GeV} [Blue-Dashed] for
[a] {\rm Im}$T$ =0.01 ${\rm TeV}^{-2}$ of $A_1^{+-}(\theta_0)$ [b]  {\rm
Re}$S$=0.01  ${\rm TeV}^{-2}$ of $A_2^{+-}(\theta_0)$. }
\label{fig:F45}
\end{figure}

Fig.~\ref{fig:F45} shows the $\theta$ integrated version
of the asymmetries plotted as a function of the cut-off angle
$\theta_0$ for $\sqrt{s}$=500 {\rm GeV} and 800 {\rm GeV}. 
Considering the $+-$ case,
we find that $A_1^{+-}(\theta_0)$ depends only on {\rm Im}$T$ and
$A_2^{+-}(\theta_0)$
depends only on {\rm Re}$S$.
For a value of 0.01 ${\rm TeV}^{-2}$ of the anomalous couplings
the asymmetries increase with the cut-off in both cases. 
This is due to the SM cross section
in the denominator $\sigma^{SM}(\theta_0)$ which decreases faster than 
the numerator.
As is clear from Fig.~\ref{fig:F45}, the asymmetries are sensitive to
the c.m. energy even in the $\theta$-integrated case.

\begin{figure}[htb]
\includegraphics[width=6.5cm,height=5 cm]{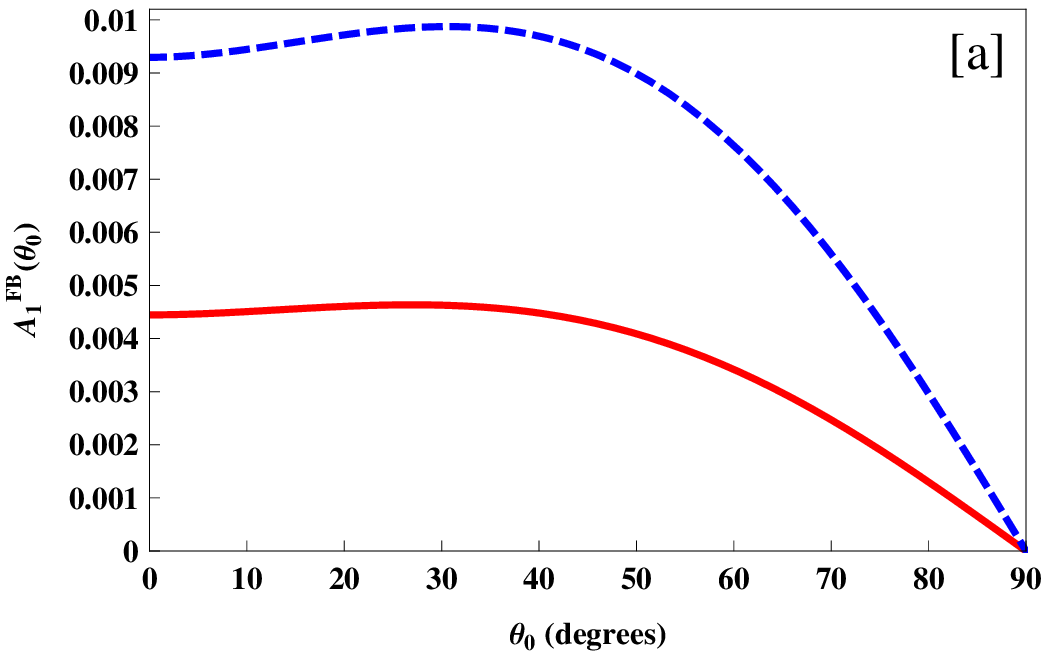}
\hspace{0.2cm}
\includegraphics[width=6.5 cm,height=5.05 cm]{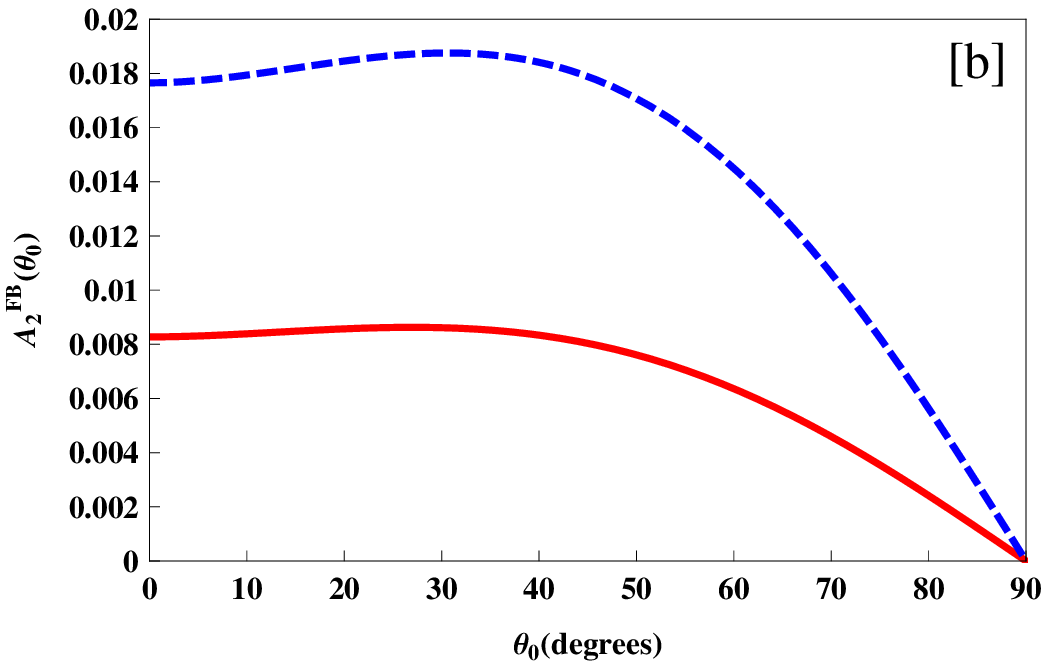}
\caption{The asymmetries as a function of $\theta_0$
for $\sqrt{s}$=500 {\rm GeV} [Red-Solid] and $\sqrt{s}$=800 {\rm GeV} [Blue-Dashed] for
[a] {\rm Im}$T$ =0.01 ${\rm TeV}^{-2}$ of $A_1^{FB}(\theta_0)$ [b]
{\rm Re}$T$=0.01  ${\rm TeV}^{-2}$ of $A_2^{FB}(\theta_0)$ for $+-$ case. }
\label{fig:F67}
\end{figure}

Since we are trying to utilise a single beam-polarization combination
for the isolation of the couplings as far as possible,
we move to the next $\theta$ integrated forward backward
asymmetry $A_1^{FB}(\theta_0)$ and  $A_2^{FB}(\theta_0)$ for the $+-$ case.
$A_1^{FB}(\theta_0)$ depends on {\rm Im}$T$ as in $A_1^{+-}(\theta_0)$ .
Fig.~\ref{fig:F67}[a] shows $A_1^{FB}(\theta_0)$ plotted as a function of cut off
for {\rm Im}$T$ =0.01 ${\rm TeV}^{-2}$. The asymmetry here is much smaller than
Fig.~\ref{fig:F45}[a] for same value of {\rm Im}$T$. This is due to the presence of
the term $\beta c^e_Ac^t_A$ before {\rm Im}$T$ in $A_1^{FB}(\theta_0)$ which is much smaller
than the term accompanying {\rm Im}$T$ in $A_1(\theta_0)$. The term in the later 
case is $(-2+3c^e_Vc^t_Vs/(s-M_Z^2))$. Fig.~\ref{fig:F67}[b] shows $A_2^{FB}(\theta_0)$
plotted for {\rm Re}$T$ =0.01 ${\rm TeV}^{-2}$. Both the asymmetries here vanishes for $\theta$=
$90\,^{\circ}$, due to the $(1-\sin^3\theta_0)$ term in the numerator.

\begin{figure}[htb]
\includegraphics[width=6.5cm,height=5 cm]{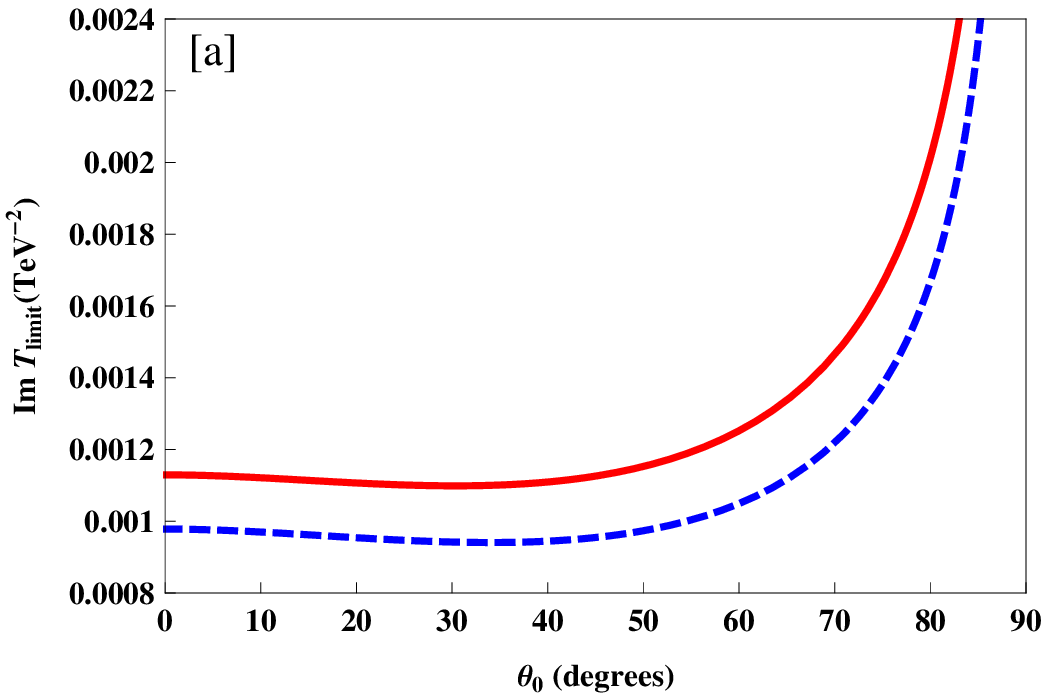}
\hspace{0.2cm}
\includegraphics[width=6.5 cm,height=5 cm]{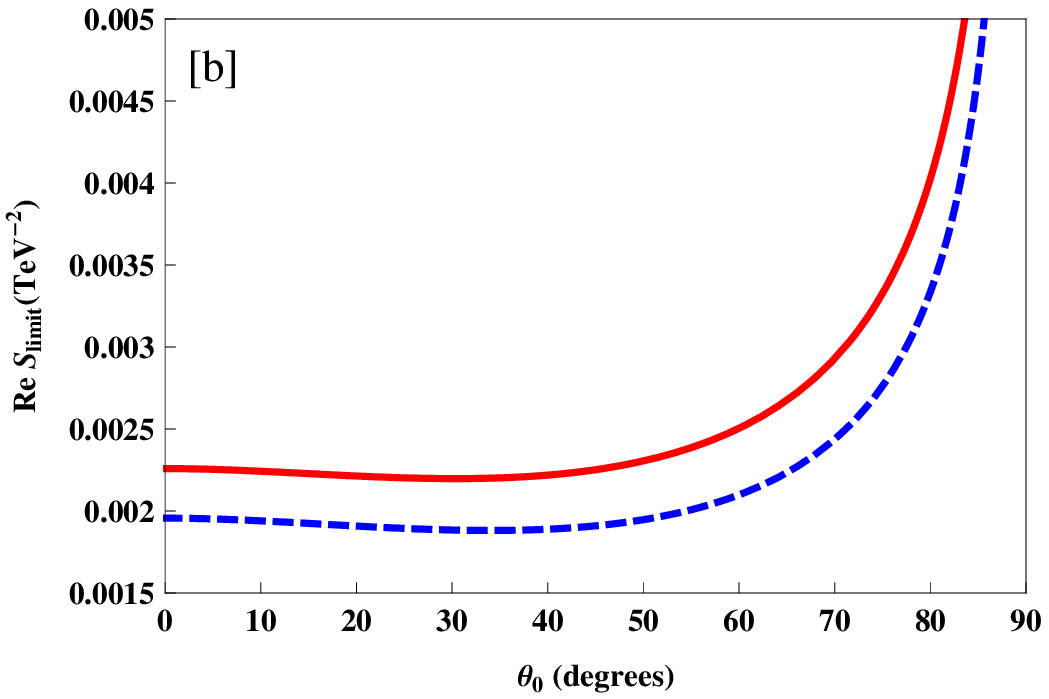}
\caption{90 \% C.L. limit obtained on [a] Im$T$ from $A_1^{+-}(\theta_0)$ [b] Re$S$ from 
$A_2^{+-}(\theta_0)$ with an integrated luminosity of 500 ${\rm fb}^{-1}$ 
at $\sqrt{s}$=500 {\rm GeV}[Red-Solid] and
$\sqrt{s}$=800 {\rm GeV}[Blue-Dashed] plotted as a function of $\theta_0$ }
\label{fig:F78}
\end{figure}

We have used the asymmetries to calculate 90 \% CL limits  
that can be obtained with ILC with an integrated luminosity
$\mathcal{L}$ of 500${\rm fb}^{-1}$, and $\sqrt{s}$=500 {\rm GeV} and 800 {\rm GeV}.
The sensitivity of the given coupling
denoted by $\mathcal{C}_{limit}$ is related to the value $A$ of the asymmetry by:
\begin{equation}
\mathcal{C}_{limit}=\frac{1.64}{|A|\sqrt{N_{SM}}}
\end{equation}
where $N_{SM}$ is the number of SM events. The coefficient 1.64 may be obtained from
statistical tables for hypothesis testing with one estimator.

Fig.~\ref{fig:F78} shows the 90\% CL limits obtained on {\rm Im}$T$ and 
{\rm Re}$S$ from $A_1^{+-}(\theta_0)$ and $A_2^{+-}(\theta_0)$ respectively.
It is seen from the Figures that the limits are relatively insensitive to
the cut off at very small values of $\theta_0$. 
The best limit is obtained for about
$\theta_0$=$40\,^{\circ}$, though any nearby value of $\theta_0$ will give similar results.
The sensitivity corresponding to {\rm Im }$T$ is 
1 $\times 10^{-3} {\rm TeV}^{-2}$ (from $A_1^{+-}(\theta_0)$)
and that corresponding to {\rm Re }$S$ is 
2 $\times 10^{-3} {\rm TeV}^{-2}$ (from $A_2^{+-}(\theta_0)$) at 
$\sqrt{s}$=800 {\rm GeV} after which 
it increases rapidly. The results for the other couplings can be obtained in
a straightforward manner. Considering $++$ and comparing it with $+-$, for
$A_1^{+-}(\theta_0)$ and $A_2^{+-}(\theta_0)$, see eqs.(\ref{eqnt1},
\ref{eqnt3}), the above sensitivities can be readily 
translated into sensitivities of other couplings. Comparing we see
{\rm Im}$S$ shares the same coefficient as 
{\rm Re}$S$, furthermore {\rm Re}$T$ and {\rm Im}$T$ also have the same coefficients. Therefore the
sensitivities in this case are the same as before {\rm i.e}.
{\rm Im}$T \leftrightarrow$ {\rm Re}$T$ and {\rm Re}$S \leftrightarrow$ {\rm Im}$S$
which are obtained by suitably interchanging the
asymmetries  $A_1(\theta_0) \leftrightarrow A_2(\theta_0)$.

\begin{figure}[htb]
\begin{center}
\includegraphics[width=6.5cm,height=5 cm]{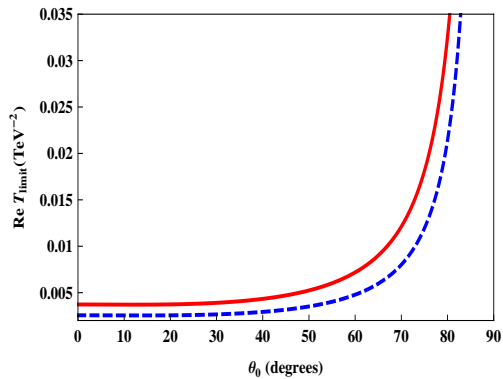}
\caption{90 \% C.L. limit obtained on  Re$T$ from $A_2^{FB}(\theta_0)$ for
$+-$ case with an integrated luminosity of 500 ${\rm fb}^{-1}$ at $\sqrt{s}$=500 {\rm GeV} [Red-Solid] and
$\sqrt{s}$=800 {\rm GeV} [Blue-Dashed] plotted as a function of
$\theta_0$.}
\label{fig:F15}
\end{center}
\end{figure}

Again returning to the fact that only $+-$ case is utilised, then $A_2^{FB}(\theta_0)$
can be used to put a limit on {\rm Re}$T$. Fig.~\ref{fig:F15} shows the behaviour pattern 
is the same as before. Here the limit obtained is about {\rm Re}$T\sim 3.5\times 10^{-3} {\rm TeV}^{-2}$.
Compared to the above results the limit obtained in this case is worse. Similarly  $A_1^{FB}(\theta_0)$
will give a limit on {\rm Im}$T\sim 7\times 10^{-3} {\rm TeV}^{-2}$,
which is worse than the previous limit obtained from $A_1^{+-}(\theta_0)$. The limits obtained
on the various couplings is summarised in Table \ref{tab:cl_limit}.

\begin{table}
\begin{center}
\begin{tabular}{||c|c|c|c|c|c|c||} \hline
&&&\multicolumn{4}{|c||}{}\\
$\sqrt{s}$ & Case &Coupling &\multicolumn{4}{|c||}{Individual limit from asymmetries}
\\ \cline{4-7}
&&&&&&\\
& & & $A_1(\theta_0)$  &$A_2(\theta_0)$   &$A_1^{FB}(\theta_0)$&$A_2^{FB}(\theta_0)$
 \\ \hline\hline
& & {\rm Re}$S$ &    &$2.3\times10^{-3} {\rm TeV}^{-2}$&  & \\ 
&$+-$ & {\rm Re}$T$ &    &  &  &$5.2\times10^{-3} {\rm TeV}^{-2}$ \\ 
500{\rm GeV} &  & {\rm Im}$T$ & $1.2\times10^{-3} {\rm TeV}^{-2}$  & &$1.0\times10^{-2} {\rm TeV}^{-2}$  & \\ \cline{2-7}
&  & {\rm Im}$S$ &$2.3\times10^{-3} {\rm TeV}^{-2}$    &  &  & \\ 
&$++$  & {\rm Re}$T$ &    &$1.2\times10^{-3} {\rm TeV}^{-2}$&  &$1.0\times10^{-2} {\rm TeV}^{-2}$ \\ 
&& {\rm Im}$T$ &   & &$5.2\times10^{-3} {\rm TeV}^{-2}$  & \\ \hline\hline
& & {\rm Re}$S$ &    &$2.0\times10^{-3} {\rm TeV}^{-2}$&  & \\ 
&$+-$ & {\rm Re}$T$ &    &  &  &$3.5\times10^{-3} {\rm TeV}^{-2}$ \\ 
800{\rm GeV} &  & {\rm Im}$T$ & $1.0\times10^{-3} {\rm TeV}^{-2}$  & &$7\times10^{-3} {\rm TeV}^{-2}$  & \\ \cline{2-7}
&  & {\rm Im}$S$ &$2.0\times10^{-3} {\rm TeV}^{-2}$    &  &  & \\ 
&$++$  & {\rm Re}$T$ &    &$1.0\times10^{-3} {\rm TeV}^{-2}$&  &$7\times10^{-3} {\rm TeV}^{-2}$ \\ 
&& {\rm Im}$T$ &   & &$3.5\times10^{-3} {\rm TeV}^{-2}$  & \\ \hline\hline
\end{tabular}
\end{center}
\caption{90 \% C.L. limit obtained on the coupling along with 
the relevant asymmetries given for the cases of 
$+-$ and $++$ case.}
\label{tab:cl_limit}
\end{table} 
 
We point out that, when the top spin is not considered,
the 90\% CL limit on the imaginary part of eq. (\ref{eqim})
is 1.6 $\times 10^{-2} {\rm TeV}^{-2}$, from \cite{BASDRPRD}. In
eq. (\ref{eqim}) keeping the value of $S$ and $T$ to be non zero 
one at a time, the limit on {\rm Im}$S$ is 1.6 $\times 10^{-2} {\rm TeV}^{-2}$
and that on {\rm Im}$T$ is 4.4 $\times 10^{-2} {\rm TeV}^{-2}$ at $\sqrt{s}$=500 {\rm GeV}, for
an integrated luminosity of 500${\rm fb}^{-1}$. 
Comparing this result with the one
obtained for $\sqrt{s}$=500 {\rm GeV}, considering the top spin, the sensitivities obtained 
are {\rm Im}$S \sim 2.3 \times 10^{-3}$  ${\rm TeV}^{-2}$ and {\rm Im}$T
\sim 1.2 \times 10^{-3}$ ${\rm TeV}^{-2}$ from $A_1^{++}(\theta_0)$
and $A_1^{+-}(\theta_0)$ respectively.
The limits in this case are an order of magnitude better than the previous one~\cite{BASDRPRD}.

%\section{Connection to polarization measurements}\label{polarization}
\section{ Selecting a sample of polarized top quarks
}\label{polarization}

In the above, we implicitly assume that it would be possible to
isolate a sample of events where the top (and anti-top) has a definite
helicity. However, in practice, this is not possible as one can only
measure polarization at a statistical level. Unlike an incoming beam of
particles, which can be prepared in a pure spin state, an outgoing
particle is not available in a pure state, but only a mixed state,
yielding only an average polarization. In order to be able to make use
of the definitions of various asymmetries which we discuss, we propose
a practical method which would serve to provide a sample with
predominantly positive or negative top helicities. This would of
course lead to a depletion of the efficiency, but would be able to 
achieve the main objective.
In the rest frame of the top, the
angular distribution of a given decay product is given 
by, 
\begin{eqnarray}
& \displaystyle
\frac{1}{\Gamma} \frac{d \Gamma}{d\cos\Theta}
 = \frac{1}{2}  (1 + P_t \kappa\cos\Theta) &
\end{eqnarray}
where $\Theta$ is the angle between the momentum of the
decay product and top spin quantization axis, which is also
the direction of the top momentum before a boost to the rest frame,
$P_t$ is the top polarization (longitudinal), and
$\kappa$ is the analyzing power for that decay channel. For a charged
lepton, $\kappa$ is 1, giving the maximum analyzing power. Thus, if 
the top rest frame is constructed event by event, and the
event then classified
depending  on whether  $\cos\Theta$ is positive or negative, we
would have two event samples, one with dominantly positive helicity
tops, and the other with dominantly negative helicity tops. The
relative sizes of the two samples will depend on the actual
polarization for that particular top (i.e. the top emitted at definite
angles $\Theta_t, \phi_t$ in the lab frame). The observables which we use
are defined with respect to $d\sigma_{NP}/d\Omega$, and the number of
events in the difference of these two samples for a particular $\Theta_t$
and $\phi_t$ (of the top) would be proportional to this $d \sigma_{NP}/d\Omega$,
though not its actual value.

To use this method completely, one has to actually generate events
including top decay, use our formulas to make predictions, and then
compare the expected number of events for a given set of anomalous
couplings with experiment and hence put a limit.
Such a procedure would give limits which are less stringent than
obtained in our analysis.
Strictly speaking we should include full spin density matrices for $
t$ (or $ \bar t$ ) production  as well as decay into a certain final
state, and consider asymmetries constructed out of the momenta of the
decay products. However, we expect that the procedure described here
will approximate such a complete description, with some reduction of
efficiency.

A full analysis including top decay entails a more complicated analysis 
with a different final state, and is beyond the scope of this work.

A similar procedure has been described in the context of $\tau$
polarization \cite{Guchait:2008ar}, where a suggestion is  made for
applying a cut on the energy fraction of the decay product of $\tau$ as
a filter for $\tau$ polarization. The same technique of applying cuts on
the energy fraction of a top decay product would be equivalent to that
of applying a cut on $\cos\Theta$ that we have suggested above.

\section{Discussion and Conclusions}\label{conclusions}

To conclude,
in this paper we have considered the process \ttbar in the presence of NP 
contributions due to scalar and tensor interactions, accounting for these 
at leading order only.  Due to chirality conservation due to the near 
masslessness of the electron, these can be manifested only in the presence 
of TP.  In contrast to earlier studies of this process, we have explicitly 
looked at the analysis of this process due to top-quark spin.  The 
immediate advantage of this even when the spin of the $\bar{t}$ is summed over 
is that it now becomes possible to disentangle the contributions of $S$ 
and $T$, which was not possible when no final state spin was measured. 
We have explicitly presented the differential cross sections, in as compact 
a manner as possible, where it is possible also to interchange the role of 
the spins of the $t$ and $\bar{t}$.

In principle, it is also possible to consider the cases where
the helicities of the top as well as the anti-top are also explicitly
resolved.  
It may be possible to carry out a study based on this,
but is beyond the scope of the present work, as the features that
we wish to study are already apparent when we sum over the helicity
of one of the other.  Furthermore, measuring both spins would lead 
to a loss in statistics thereby making this option less attractive.

We have then carried out an extensive numerical analysis based on these 
cross sections by defining suitable integrated asymmetries.  By employing 
realistic integrated luminosity we have obtained 90 \% C.L. limits that can be 
placed on the NP couplings. With an integrated luminosity
of 500 ${\rm fb}^{-1}$ and realistic beam polarizations the limits on 
real and imaginary parts for $T$ and 
$S$ are of order $ 10^{-3} {\rm TeV}^{-2}$
at $\sqrt{s}$=500 and 800 {\rm GeV}. These limits are found to be better
by an order of magnitude compared to the previous case. These thus fare better than the corresponding 
analysis based on only momentum measurements when the spins of both final 
state particles are summed over.  It is also of interest to compare
these numbers with the naturalness constraints $O(10^{-3}) {\rm TeV{^{-2}}}$ 
on Re $T$ from the $g-2$ of the electron, which is the most
stringent one, whereas weaker constraints exist on the corresponding
imaginary part from the electron electric dipole moment, see 
ref.~\cite{BASDRPRD}.

We have assumed in this work
perfect beam polarization.  If we were to 
take as in ref.~\cite{BASDRPRD} $P_e=0.8$ and $P_{\bar e}=-0.7$, 
then once again we would lose a factor of $\approx 0.7$ in the asymmetry
with a corresponding lowering of sensitivity.
It must, however, be mentioned that
one cannot directly isolate events with top helicities of +1 or -1. Hence
to measure the asymmetries we discuss, one would have to carry out a
subtraction of events in two kinematic regions of the decay products
corresponding to positive and negative polarizations of the top. Doing
so would entail a loss of efficiency to a certain extent. We have not
taken this into account.

In order to understand the nature of the spin-momentum and spin-spin 
correlations, we have made contact with the general inclusive formalism 
developed in refs.~\cite{BASDR1,BASDR2}.  This has required us to 
explicitly spell out the spin vector for the top quark and to identify the 
spin structure functions.  Interestingly {\rm Re}$S$ induces only one type of 
spin structure function, while {\rm Re}$T$ induces three types of spin 
structure functions.  Analogous statements hold for the imaginary parts as 
well.  The advantage of this formalism is that one is able to explicitly 
study the properties of the correlations under the discrete symmetries C, 
P and T.  Our discussion is more explicit than the discussion in the
context of the inclusive process in ref.~\cite{DR2}.   It must be
emphasized that to comprehend
the structure of the spin-momentum correlations and the discrete symmetry
properties of each of the terms in the distributions without this
framework would be nearly impossible.  
This discussion is presented in the Appendix.  

Finally, it
must also be mentioned here that the process under consideration is of 
interest in the context of electroweak Sudakov processes with TP, see 
e.g. ref.~\cite{hep-ph/0311260}.
It would be interesting to actually carry out a study 
to see how these effects could mimic effects arising from NP of this type. 
This could be the topic of a future study.

\bigskip
{\bf Acknowledgements:}  BA thanks the Homi Bhabha Fellowships Council
for support through an award. SDR thanks the Theory Group of TIFR
Mumbai, where part of this work was done, for hospitality.

\bigskip

\appendix*
\section{Interpretation in the general inclusive framework}\label{inclusive}

As mentioned in the Introduction one of the main reasons for
considering TP for at least one of the beams is that NP
of the $S$ and $T$ type will not otherwise appear in distributions at linear
order.  This feature a result of chirality conservation in the
limit of massless electrons, is also the cornerstone of the
analysis for a general inclusive process recently considered in the context of
the ILC in refs.~\cite{BASDR1,BASDR2}.  In spirit this approach
of retaining NP at linear order follows the one
proposed in the context of neutral currents by
Dass and Ross (DR)~\cite{DR1,DR2}.  

The outcome of this approach is that the spin-momentum
correlations involving the incoming particles and the
momentum of the observed particle, or the momenta of
the two observed particles uniquely fingerprints the
Lorentz structure of the NP.  
The approach here is fruitful in many ways. 
For instance, it was concluded that no CP violating 
couplings of the type V and A would show up in the inclusive process 
$e^{+} e^{-}\rightarrow h(p) X$ via spin-momentum correlations
if NP amplitude were to interfere with QED contribution to
the SM amplitude.  
This conclusion also remained true for the $Z-$ contribution
to the SM amplitude as well, which was explicitly
demonstrated in ref.~\cite{BASDR1}.In processes such as the above,
the physics would be described entirely in terms of `momentum
structure functions'.
Note that whereas in ref.~\cite{DR2}
the spin and momentum of one observed particle was also
considered, this has not been done in the context of ILC physics.
The general inclusive framework is described in terms of
`structure functions' associated with the inclusive final state,
one for each type of interaction as well as for the various
vectors from which Lorentz invariant amplitudes were constructed,
when contracted with the leptonic tensor built out of the 
interference of the SM and NP diagrams, essential features
of which will be recalled below.

Here we provide a concrete illustration for the case
of the observed particle being the top quark, {\it viz.} a spin-1/2 particle,
with the task at hand now being the identification of what will
be called the `spin structure functions'.
The objective now is to relate the general inclusive framework
to that of the computed distributions for the explicit $t\bar{t}$ final state
which explicitly receive contributions that are spin-independent and
those that are not, and the latter expressed in a straightforward manner on
$h$.

The correspondence to the framework for the helicity independent
part of the correlations that appears in the
process represented by eq.(\ref{process1}) has already been done and
presented in~\cite{BASDR1}.  
This correspondence
was straightforward and no detailed discussion was presented.
For the helicity dependent part, however, the correspondence
is more involved and it is worth presenting a detailed
discussion. 

\subsection{Formalism for the spin-momentum correlations}

To begin the discussion, we begin by
observing that the spin-momentum correlations amongst
those of the incoming particles and the outgoing particles will
arise from the interference between the SM currents with the NP
`currents' which requires us to consider the trace:
 \begin{equation}\label{trace}
{\rm Tr}[(1-\gamma_5 h_{+} + \gamma_5 \slashed s_{+})\slashed p_{+}\gamma_\mu(g_V^e-g_A^e \gamma_5)
(1+\gamma_5 h_-+\gamma_5 \slashed s_-)\slashed p_-\Gamma_i]H^{i\mu }.
\end{equation}
following the notation of refs.~\cite{BASDR1,BASDR2},
where $i$ now is a generic index that denotes the scalar, pseudoscalar
and tensor interactions, $h_\pm$ are the degrees of longitudinal
polarization and $s_{\pm}$ represent the transverse
polarizations of the positron and electron.
In terms of the scalar, pseudoscalar and tensor couplings
$g_S, g_P$ and $g_T$ of the electron, and structure functions
describing the inclusive process given by $F^r$, $F_1^{rut}, F_2^{r},
PF_1^{rut}$ and $PF_2^r$, we may express the vertices
$\Gamma_i$ and $H^{i\mu }$ as
\begin{equation} 
\Gamma =g_s+i g_p\gamma_5
\end{equation}
and
\begin{equation}
H^{S}_{\mu}=(r_{\mu}-q_{\mu}\frac{r \cdot q}{q^2})F^{r}
\end{equation}
$r$ is $p_t$, $s_t$ or $ n(n_{\mu}\equiv 
\epsilon_{\mu\alpha\beta\gamma}p_t^\alpha s_t^\beta q^\gamma)$,
for $S$ and $P$ type NP interactions, and
\begin{equation}
\Gamma_{\rho\tau}=g_T\sigma_{\rho\tau}.
\end{equation}
\begin{eqnarray}
H^{T}_{\mu\rho\tau} &=& (r_\rho u_\tau-r_\tau u_\rho)t_\mu F_1^{rut}+(g_{\rho\mu}r_\tau-g_{\tau\mu} r_{\rho})F_2^r \nonumber \\
                    && + \epsilon_{\rho\tau\alpha\beta}r^\alpha u^{\beta} t_\mu PF_1^{rut}+\epsilon_{\rho\tau\mu\alpha}r^\alpha PF_2^r
\end{eqnarray}
where $r$ is $q$, $p_t$ or $s_t$. Similarly $u$ is chosen to be one
of $q$, $p_t$, $s_t$ or $n$ and $t$ being $p_t$, $s_t$ or $n$ respectively
for $T$ type interactions.
\footnote{The structure functions for the case of the process
given by eq.(\ref{process1}) appear with no superscripts
as there is only one vector $p$, the momentum of
the observed particle, on hand.  By straightforward
inspection it was inferred in ref.~\cite{BASDR1} that
a correspondence between the inclusive process and
\ttbar production could be inferred:
the correspondence was given
as the structure functions
${\rm Re} (g_P F)$ and ${\rm Re} (g_T F_2)$ as arising from
${\rm Im} S_{RR}$ and ${\rm Im} T_{RR}$.}
One may then evaluate the spin-momentum correlations
due to the various structure functions in a straightforward
manner.  For ease of comparison, one may also compare against
spin-momentum correlations that are tabulated in the
appropriate tables in ref.~\cite{BASDR2}. In terms of the
kinematic quantities suitable to the process,
which are $\vec{K}\equiv (\vec{p}_--\vec{p}_+)/2=E\hat{z}$,
$q\equiv p_-+p_+$ with $q^0=2 E$ and $\vec{q}=0$.
The analysis may be readily extended to the case of
spin-momentum and spin-spin (helicity) correlations, where the latter
is that of the observed final state particle.

In order to achieve this end, we first require an explicit
representation for the vector describing the spin of the
observed particle, $s_t$.  In the helicity eigenbasis,
the components of the spin vector $s_t$ are
(see, e.g., eq. (3.155) in ref.~\cite{Greiner}) 
\begin{eqnarray}
&
\left({|\vec{p}_t|}/{m_t},E\,\vec{p}_t/(m_t\, |\vec{p}_t|)\right). &
\end{eqnarray}
This representation for the spin vector in the helicity basis follows from 
considering free spinors for the quark and anti-quark, first in their 
respective rest frames and then boosting them to the laboratory frame.  By 
considering covariant generalization of the appropriate Pauli matrices to 
define the spin and introducing spin projection operator and choosing the 
spin direction to be that of the momentum, which is the appropriate choice 
for the helicity basis we obtain the desired expression.  
With this explicit representation and with the identification of
the vector $\vec{r}$ as $\vec{s}_t$ one may now turn to
the appropriate tables in ref.~\cite{BASDR2}.  

Consider now the case
of $++$ (we introduce a notation $ij,\, i,j=+,-$ to denote  
the sense of the polarization of the electron ($i$) and positron ($j$) 
respectively).  We may now compute the correlation directly, or we
need simply to look at the correlation due to the structure function
Im$(g_s F^{s_t})$ which from Table 1 of ref.~\cite{BASDR2} reads:
\begin{eqnarray}
& 2 E g_V^e \vec{K}\cdot \left(\vec{s}_++\vec{s}_-\right)\times \vec{s}_t. & 
\end{eqnarray}
The above evaluates in terms of the kinematics of the reaction
at hand to
\begin{eqnarray}
& \frac{2 E^3}{m_t} g_V^e \sin\theta \sin\phi. &
\end{eqnarray}
Now looking into the explicit expression for the distribution
for $t\bar{t}$ production, eq. (\ref{diffcspp1}), one may readily see that 
the same angular dependence of this correlation is the one that
accompanies Im$S$.
Stated differently, the four-fermion contact interaction
due to Im$S$ induces the structure function Im$(g_s F^{s_t})$.
This completes the first correspondence that we are after.

This analysis may now be extended to the four-fermion
contact interactions due to $T$.  This interaction
induces more than one kind of structure function, which
can be explicitly obtained.
It turns out that there are 
three such structure functions,  which are presented 
with the respective correlations read off from Table 3 of
ref.~\cite{BASDR2}, and the corresponding expressions  
in terms of the present kinematics as
well as the four-fermion interactions which are responsible
for inducing the relevant structure functions
are shown in Table~\ref{tab:sig_f1}.

\newcommand\T{\rule{0pt}{4.0ex}}
\newcommand\B{\rule[-1.2ex]{0pt}{0pt}}
\begin{table}
\begin{center}
\begin{tabular}{||c|c|c|c||}\hline
&\multicolumn{2}{|c|}{}&\\
Structure function &\multicolumn{2}{|c|}{Correlation}&Coupling
\\ \cline{2-3}
&&&\\
 & Vector form & Polar form & 
\\  \hline\hline
Re$(g_TPF_2^{s_t})$ & $\T  4 E^2 \vec{s}_t\cdot (\vec{s}_++\vec{s}_-)$ &
	$\frac{4 E^3}{m_t} g^e_V \sin\theta \cos\phi \B $ & Re $T$ \\  \hline\hline
Re$(g_T F_1^{p_tqs_t})$ & $\T 4E(E[p_t^0(\vec{q}\times\vec{s}_t^T)-q^0
		(\vec{p}_t\times\vec{s}_t^T]\cdot$ & $\frac{8 E^4}{m_t}
		|\vec{p}_t|g_V^e \sin\theta\cos\theta\sin\phi$ & Im $T$ \\
& $(\vec{s}_++\vec{s}_-)+[(\vec{p}_t\times\vec{q})\cdot\vec{p}_+]\vec{s}_t\cdot
			(\vec{s}_+-\vec{s}_-)) \B$ & & \\   \hline \hline
Im$(g_T F_1^{p_tqs_t})$ & $\T  4E^2([(\vec{p}_t^T\cdot\vec{s}_t^T)\vec{q}-
		(\vec{q}^T \cdot \vec{s}_t^T)\vec{p}_t]\cdot$ & $\frac{8 E^4}{m_t}
		|\vec{p}_t|g_A^e \sin\theta\cos\theta\cos\phi$ & Re $T$ \\
& $(\vec{s}_+-\vec{s}_-)+(q^0p_t^3-q^3p_t^0)\vec{s}_t\cdot
			(\vec{s}_+-\vec{s}_-)) \B$ & & \\ \hline 
\end{tabular}
\end{center}
\caption{Structure functions along with the correlation
in vector and polar form for the $+ +$ case and coupling which
give rise to the structure functions.}
\label{tab:sig_f1}
\end{table}

Turning now to the case of $+-$, we now consider the 
analogous correlations.
In case of the scalar the structure function
Im$(g_p f^{s_t})$ generates 
(see Table 1 of ref.~\cite{BASDR2}) the correlations
\begin{eqnarray}
-2E^2(\vec{s}_+-\vec{s}_-)\cdot \vec{s}_t
\end{eqnarray}
which evaluates to 
\begin{eqnarray}
-\frac{2E^3}{m_t}g^e_V\sin\theta\cos\phi.
\end{eqnarray}
It may be readily seen by inspecting eq.(\ref{diffcspp2})
this structure function is generated by the four-fermion
interaction due to Re$S$. 

The tensor part gives rise to three structure functions
which are presented in Table ~\ref{tab:sig_f2}. 
The corresponding correlations are read off from
Table 3 of Ref.~\cite{BASDR2}, and the explicit
representation in terms of the kinematics of the
present process are also tabulated, as well
as the NP terms that induce these structure functions.

\begin{table}
\begin{center}
\begin{tabular}{||c|c|c|c||}\hline
&\multicolumn{2}{|c|}{}&\\
Structure function &\multicolumn{2}{|c|}{Correlation}&Coupling
\\ \cline{2-3}
&&&\\
 & Vector form & Polar form &
  \\  \hline\hline
Re$(g_TF_2^{s_t})$ & $ \T -4E(\vec{s}_+-\vec{s}_-)\times\vec{K}\cdot\vec{s}_t$ &
	$-\frac{4 E^3}{m_t} g^e_V \sin\theta \sin\phi \B$ & Im$T$ \\  \hline\hline
Re$(g_T PF_1^{p_tqs_t})$ & $ \T 4E^2([(\vec{p}_t^T\cdot\vec{s}_t^T)\vec{q}-
		(\vec{q}^T \cdot \vec{s}_t^T)\vec{p}_t]\cdot$ & $-\frac{8 E^4}{m_t}
		|\vec{p}_t|g_V^e \sin\theta\cos\theta\cos\phi$ & Re$T$ \\
& $(\vec{s}_++\vec{s}_-)+(q^3p_t^0-q^0p_t^3)\vec{s}_t\cdot
			(\vec{s}_+-\vec{s}_-)) \B$ & & \\  \hline \hline
Im$(g_T PF_1^{p_tqs_t})$ &$ \T 4E(E[p_t^0(\vec{q}\times\vec{s}_t^T)-q^0
		(\vec{p}_t\times\vec{s}_t^T]\cdot$ & $\frac{8 E^4}{m_t}
		|\vec{p}_t|g_A^e \sin\theta\cos\theta\sin\phi$ & Im$T$ \\
& $(\vec{s}_+-\vec{s}_-)-[(\vec{q}\times\vec{p}_t)\cdot\vec{p}_+]\vec{s}_t\cdot
			(\vec{s}_++\vec{s}_-)) \B$ & & \\ \hline 
\end{tabular}
\end{center}
\caption{Structure functions along with the correlation in vector and polar form
for the $+ -$ case and coupling 
which give rise to the structure functions.}
\label{tab:sig_f2}
\end{table} 

In summary, we have presented here in detail the generalization of
the result in ref.~\cite{BASDR1} for the momentum structure function
to the spin structure functions induced by the four-Fermi interactions.
These results are helpful in understanding the C (charge conjugation), 
P (parity) and T (time-reversal) properties
of the correlations which is the subject of the forthcoming subsection.

\subsection{Properties under C, P and T}
A discussion on the properties of the correlations under the
discrete symmetries of C, P and T  
appearing in the distribution is the subject of this
subsection.  We note here that T will represent
na\"ive time reversal i.e. reversal
of all spins and momenta, without interchange of initial and final
states.  We will see that 
the helicity dependent part of the correlations
are substantially richer in structure compared to their
helicity independent counterparts.  

We begin by noting that 
the differential cross sections corresponding to antiparallel
or parallel $e^-$ and $e^+$ polarization have both CP-odd as well as CP-even
quantities compared to the helicity independent parts.
The additional features arise from the $h$ and $z'$ 
dependent quantities.  

Let us keep in mind that at the level of the effective Lagrangian,
if the projection operators are expanded out completely,
and if the real and imaginary parts of $S$ and $T$ are 
separated, it can be checked that terms occurring with Re$S$ and Re$T$ are
CP even, whereas the ones with Im$S$ and Im$T$ are CP odd.
The question one may then ask is how this can be seen in
the individual terms appearing in the distributions.

In order to achieve this end, 
the terms in the cross sections for the various spin configurations
of the electron and positron spins have to be written
in terms of the  momentum and spin correlations which are explicitly
even or odd.  The 
requisite combinations are presented in Table ~\ref{tab:sig_f}.

\begin{table}
\begin{center}
\begin{tabular}{||c|c|c|c|c|c||} \hline
&&\multicolumn{2}{|c|}{}&&\\
 Case &NP Term &\multicolumn{2}{|c|}{Correlation}&CP&T
\\ \cline{3-4}
&&&&&\\
 & & Polar form  & Vector form  &&
 \\ \hline\hline

++/-- --  & Re$T$ &$\sin\theta\cos\phi$  
 \T &${\displaystyle\frac{(\vec s_++\vec s_-)\cdot(\vec p_t-\vec p_{\bar{t}})}{|\vec p_t-\vec p_{\bar{t}}|\B}}$ 
 &+ &+        \\ 
    & Im$S$ &$\sin\theta\sin\phi$   
 \T &${\displaystyle\frac{(\vec p_- - \vec p_+)\times ((\vec s_++\vec s_-)\cdot
(\vec p_t-\vec p_{\bar{t}}))}{|\vec p_- - \vec p_+||\vec p_t-\vec p_{\bar{t}}|\B}}$ 
&+ &--            \\  \hline\hline
+ --/-- +   & Re$S$  &$\sin\theta\cos\phi$  
\T &${\displaystyle\frac{(\vec s_+-\vec s_-)\cdot(\vec p_t-\vec p_{\bar{t}})}{|\vec p_t-\vec p_{\bar{t}}|\B}}$ 
 &-- &+         \\ 
 & Im$T$   &$\sin\theta\sin\phi$   
\T &${\displaystyle\frac{(\vec p_- - \vec p_+)\times ((\vec s_+-\vec s_-)\cdot
(\vec p_t-\vec p_{\bar{t}}))}{|\vec p_- - \vec p_+||\vec p_t-\vec p_{\bar{t}}|\B}}$ 
&-- &--            \\  \hline 
\end{tabular}
\end{center}
\caption{CP and T properties of different correlations appearing in the
cross section for the reaction \\
$e^{-}(p_-,s_-)+e^{+}(p_+,s_+)\rightarrow t(p_t,s_t)+\bar{t}(p_{\bar{t}},
s_{\bar{t}})$.  }
\label{tab:sig_f}
\end{table} 

The entries in Table \ref{tab:sig_f}
enable us to decipher the CP properties in the following manner.
Let us first note that the quantities $h$ and $hz'$ may
be defined in the following manner: 
\begin{eqnarray}
& h ={\displaystyle\frac{(\vec s_t+\vec s_{\bar{t}})
\cdot(\vec p_t-\vec p_{\bar{t}})}{|\vec p_t-\vec p_{\bar{t}}|}}, & \\ 
& hz' = {\displaystyle\frac{(\vec s_t-\vec s_{\bar{t}})
\cdot(\vec p_t-\vec p_{\bar{t}})}{|\vec p_t-\vec p_{\bar{t}}|}}. 
&   
\end{eqnarray}

Thus we may explicitly see that CP$(h)=+$,  T$(h)=+$
and CP$(hz')=-$,  T$(hz')=+$.
From the expressions for the distributions, it can be checked considering the
CP and T properties from the table, that the terms coming with Re$S$ and
Re$T$ and those coming with Im$S$ and Im$T$ 
are both CPT even.  This is due to
the fact that the effective Lagrangian is Hermitian.
As there are no non-Hermitian terms, there are no CPT odd terms,
see ref.~\cite{hep-ph/9411398}.

Let us now consider the implications of this for the entries
in Table \ref{tab:sig_f} with an explicit example.
Consider the correlation in the distribution for the $++$ case
accompanying Im$S$, which appears in
the combination $hz' \sin\theta\sin\phi$, see eq.(\ref{diffcspp1}).
It may now be readily seen from Table \ref{tab:sig_f}
that the term $\sin\theta\sin\phi$ CP even and T odd, whereas
$hz'$ is CP odd and T even,  and
as a result the entire term is CP odd and CPT even.
Analogous exercises may be carried for all the correlations
appearing in the explicit differential cross sections.

Let us again emphasize that the top-helicity analysis allows us to
isolate the $T$ and $S$ contributions for the following reasons:
the four-Fermi interaction due to $T$ gives rise to three
different spin structure functions each with its characteristic
spin-momentum correlation, whereas the one due to S gives rise
to only one spin structure function.  This is in contrast to the
measurement with no top spin analysis, where both $T$ and $S$ give rise
to only one momentum structure function each, and also give rise to
the same spin-momentum correlation, which is why
we are unable to disentangle the contribution in this case.  This
gives us to an explicit understanding of these spin-momentum correlations
which one could not have obtained by merely inspecting the distributions.
Secondly, without the present considerations we would not have been
able to discuss the 
C, P and T properties of the correlations.

Thus the general inclusive framework and the structure of the 
spin-momentum and spin-spin correlations provide a useful guide
for understanding the properties of the correlations obtained in
our exclusive process.

\end{document}